\documentclass[preprint,aps]{revtex4}

\usepackage{graphicx}
\usepackage{dcolumn}
\usepackage{bm}

\usepackage{epsf}
\usepackage{anysize}
\usepackage{graphicx}
\marginsize{1 in}{1 in}{1in}{1.40in}

\usepackage{graphicx}
\usepackage{dcolumn}
\usepackage{bm}
\usepackage{amsmath}
\usepackage{amssymb}
\usepackage{mathrsfs}
\usepackage[latin1]{inputenc}
\usepackage{amsmath}

\makeatletter
  \newcommand\figcaption{\def\@captype{figure}\caption}
  \newcommand\tabcaption{\def\@captype{table}\caption}
\makeatother

\newcommand{\dd}{{\rm d}}

\newcommand{\sd}{Schr\"{o}dinger }

\newcommand{\lmd}{\lambda }

\newcommand{\G}{{\rm{Stab}}}
\newcommand{\J}{\mathcal{J}}
\newcommand{\U}{\mathcal{U}}
\newcommand{\n}{\mathcal{N}}
\newcommand{\hil}{\mathcal{H}}
\newcommand{\h}{{\rm Hess}}

\newcommand{\R}{{\mathbb{R}}}
\newcommand{\tr}{{\rm Tr}}

\newcommand{\Sp}{{\rm Sp}}
\newcommand{\OSp}{{\rm OSp}}
\newcommand{\ISp}{{\rm ISp}}
\newcommand{\spl}{{\rm sp}}
\newcommand{\SUM}{{\rm SUM}}

\begin{document}


\title{Optimal Control Theory for Continuous Variable Quantum Gates}

\author{Rebing Wu}
\altaffiliation{Department of Chemistry, Princeton University,
Princeton, New Jersey 08544, USA}


\author{Raj Chakrabarti}
\altaffiliation{Department of Chemistry, Princeton University,
Princeton, New Jersey 08544, USA}

\author{Herschel Rabitz}
\altaffiliation{Department of Chemistry, Princeton University,
Princeton, New Jersey 08544, USA}

\begin{abstract}

We apply the methodology of optimal control theory to the problem of
implementing quantum gates in continuous variable systems with
quadratic Hamiltonians. We demonstrate that it is possible to define
a fidelity measure for continuous variable (CV) gate optimization
that is devoid of traps, such that the search for optimal control
fields using local algorithms will not be hindered. The optimal
control of several quantum computing gates, as well as that of
algorithms composed of these primitives, is investigated using
several typical physical models and compared for discrete and
continuous quantum systems. Numerical simulations indicate that the
optimization of generic CV quantum gates is inherently more
expensive than that of generic discrete variable quantum gates, and
that the exact-time controllability of CV systems plays an important
role in determining the maximum achievable gate fidelity.  The
resulting optimal control fields typically display more complicated
Fourier spectra that suggest a richer variety of possible control
mechanisms. Moreover, the ability to control interactions between
qunits is important for delimiting the total control fluence. The
comparative ability of current experimental protocols to implement
such time-dependent controls may help determine which physical
incarnations of CV quantum information processing will be the
easiest to implement with optimal fidelity.
\end{abstract}

\maketitle

\section{INTRODUCTION} The optimal control of
quantum dynamical systems has become a subject of intense interest
in chemistry, physics and most recently, information theory
\cite{PeiDah1988,ShaBru2006}. Over the past several years, it has
become clear that the physical implementation of logical gates in
quantum information processing (QIP) may be facilitated by using the
methods of optimal control theory (OCT)
\cite{PalKos2002,TesKur2001,Grace2006a,Khaneja2001,Khaneja2002a,Khaneja2002b}.
When implementing a quantum logic gate through OCT, the distance
between the real and ideal quantum unitary transformation can be
used as a metric for assessing optimality \cite{PalKos2002}. The map
between admissible controls and associated values of this distance
is referred to as the control landscape; extremal control solutions
to the problem correspond to critical points of this map. These
landscapes were recently shown to universally possess very simple
critical topologies for finite-dimensional quantum gates, with no
suboptimal traps impeding optimal searches, irrespective of the
system Hamiltonian for controllable systems
\cite{RabMik2005,Mike2006b}.

Prior work on the implementation of quantum gates using OCT has been
directed toward discrete QIP, as hypothetically carried out on the
so-called quantum spin computers originally discussed by Benioff and
Feynman \cite{Benioff1982,Feynman1985}. These computers, in which
information is carried as quantum bits (qubits) \cite{NieChu2000}
encoded in discrete systems like electron spins or two-level atoms,
are the quantum version of digital classical computers. Classical
information can be carried by either a discrete (digital) signal or
a continuous (analog) signal. Quantum information can also be
carried by continuous (infinite-dimensional) systems, such as a
harmonic oscillator, rotor, or modes of the electromagnetic field
\cite{LloBra1998}. Quantum information processing over continuous
variables (CV) can be thought of as the systematic creation and
manipulation of continuous quantum bits, or qunits
\cite{LloBra1998}.

Importantly, CV QIP may be less susceptible to drift than its
classical counterpart. Cleverly encoded quantum states can be
restandardized and protected from the gradual accumulation of
small errors, or from the destructive effects of decoherence
\cite{Shor1995}. Moreover, compared to discrete QIP,
continuous-variable QIP has several practical advantages,
originating for example in the high bandwidth of continuous
degrees of freedom, that have spurred substantial interest in its
generic properties. Significant advances have recently been made
in the experimental implementation of continuous QIP, including
the demonstration of quantum teleportation over continuous
variables \cite{Fur1998}. As experimental methodologies improve,
it becomes important to consider how such implementations could be
enhanced through the systematic application of OCT.

Early control studies on continuous quantum systems focused on the
manipulation of quantum scattering states in bond-selective
control (i.e., dissociation or association of atoms) of molecular
systems \cite{ZhaoRice1991,ShaBru2005}. Such systems were shown to
be associated with dynamical symmetries represented by noncompact
Lie groups with infinite-dimensional unitary representations
\cite{WuTarn2006}. A criterion for approximate strong
controllability was given, showing that such systems, which
possess an uncountable number of levels, can be well manipulated
using a finite number of control fields.

Here, we examine OCT problems pertaining to an important class of
continuous quantum gates, namely those that can be represented as
symplectic transformations of the quadrature vectors in the
Heisenberg picture. This gate set is referred to as the Clifford
group \cite{BarSan2002}.  Although universal quantum computation
over continuous variables requires higher-order nonlinear
operational gates, networks using only Clifford group gates have
numerous important applications in the area of quantum
communication, and in fact, these gates are in many ways easier to
implement over continuous variables than over discrete variables
\cite{BraunPati2003}. For example, quantum error correcting codes,
which are essential for overcoming the effects of errors and
decoherence, use only Clifford group gates for encoding and
decoding. Other important protocols in QIP, like quantum
teleportation, also rely solely on Clifford group gates and
related measurements. The Clifford group gates are sufficient to
represent any CV quantum computation that can be efficiently
simulated on a classical computer. As such, the symplectic gate
formalism also has applications in the context of reversible
analog classical computation.

In this paper, we investigate the implementation of CV quantum
gates via OCT. We carry out this analysis in two stages. First, we
analyze the topology of the optimal control landscape for
symplectic gate fidelity upon the assumption of full
controllability of the underlying control systems, which is
demonstrated to be free of local traps that might otherwise impede
the optimization process. This analysis is Hamiltonian-independent
except for the assumption of full controllability and therefore
the conclusions are generic for CV quantum computation. Next, we
carry out numerical OCT calculations for various CV gates,
comparing to the analogous discrete variable gates with two
physical models. We identify characteristic differences between
these two problems both in terms of control optimization
efficiency and the complexity of the associated optimal
Hamiltonians.

The paper is organized as follows. Section II summarizes the
symplectic geometry arising in quantum linear optics and provides
preliminaries on the symplectic gates applied in CV QIP. Section
III analyzes the landscape topology for symplectic gate fidelity
and its restriction on the maximal compact subgroup of the
symplectic group. {In Section IV, we discuss several physical
models for CV QIP and compare their properties as control
systems.} In Section V, we carry out OCT calculations for specific
CV gates and algorithms using these model Hamiltonians, comparing
with the corresponding problems for discrete quantum gates.
Finally, Section VI draws general conclusions regarding OCT for CV
quantum gates versus discrete quantum gates.

\section{SYMPLECTIC GEOMETRY AND SYMPLECTIC GATES}

Formulation of optimal control theory for CV gates is simplified
by framing the time evolution of the canonical observable
operators in terms of group theory \cite{Arnold1989}. For example,
suppose that the system is be realized as a quantized
electromagnetic field. Let $\hat a_i$ and $\hat a_i^\dag$
represent the creation and annihilation operators corresponding to
the $i$-th mode of the field. These operators are related to the
position and momentum operators by
$$\hat q_i=\frac{1}{\sqrt{2}}(\hat a_i+\hat a_i^\dag),\quad \hat p_i=\frac{1}{\sqrt{2}}(\hat a_i-\hat a_i^\dag)$$
Then the time evolution of the vector of quadratures is associated
with a time evolution of the $2N$-dimensional vector of creation
and annihilation operators $\xi=(\hat a_1,\cdots,\hat a_N, \hat
a_1^{\dag}, \cdots, \hat a_N^{\dag})$.

The system Hamiltonian $\hat H(t)$ generates a one-parameter
family of evolution transformations $U(t)$ on the Hilbert space
$\hil$ that obeys the \sd equation
\begin{equation}\label{U-sd}
\frac{\partial {U}(t)}{\partial t}=-
\frac{i}{\hbar}~\hat{H}(t){U}(t),
\end{equation}
where the parameter is the time, and $\hat H$ is assumed to be a
quadratic Hamiltonian that takes on various forms depending on the
nature of the coupling between the oscillator modes
\cite{ArvMuk1995}. Denote the vector of quadratures of quantum
observables $\hat z=(\hat q_1,\cdots, \hat q_N;\hat
p_1,\cdots,\hat p_N)^T$. The evolution propagator transforms the
quadrature vectors linearly through
$$U:~~\hat z_\alpha~\rightarrow ~U^\dagger(t) \hat z_\alpha U(t)=\sum_\beta S_{\alpha\beta}(t)\hat z_\beta,$$
where the matrix $S(t)$ is an element of the symplectic group
$\Sp(2N,\R)$. The symplectic group is defined as the set of
$2N\times 2N$ matrices that satisfies $S^T J S=J$, where
$$J=\left(%
\begin{array}{cc}
   & I_N \\
  -I_N &  \\
\end{array}%
\right).$$ Thus, the matrix $S$ captures the Heisenberg equations
of motion for the operators $\hat z_i$, and the unitary propagator
$U$ forms the metaplectic unitary representation of $S$ in
$\Sp(2N,\R)$. Correspondingly, let $H=\{h_{ij}\}=H^T$ be the
matrix representation of the Hamiltonian, i.e., $\hat
H(t)=\sum_{i,j}h_{ij}(t)\hat z_i \hat z_j$, which belongs to the
Lie algebra $\spl(2N,\R)$ of $\Sp(2N,\R)=\{JH~|~H^T=H\}$
\cite{DraNer1988} with dimension being $N(2N+1)$. Through this
representation, the symplectic matrix $S(t)$ associated with
$U(t)$ follows a classical Hamiltonian evolution equation
\begin{equation}\label{S-sd}
\frac{\dd {S}(t)}{\dd t}=JH(t)S(t).
\end{equation}
The infinite dimensional unitary operator $U(t)$ in (\ref{U-sd})
carries the (metaplectic) unitary representation of these
symplectic transformations on the Hilbert space $\hil$ of quantum
states in the \sd picture \cite{ArvMuk1995}.

Another important class of transformations are the displacements
that shift $\hat q$ and $\hat p$ by constants:
$$\hat q \rightarrow \hat q+a, \quad \hat p \rightarrow \hat
p+b.$$ The combination of displacements with homogeneous
symplectic transformations forms the inhomogeneous or
affine symplectic group $\ISp(2N,\R)$ in the form of $$S_c=\left(%
\begin{array}{cc}
  S & c \\
  0 & 1 \\
\end{array}%
\right),~~~~S\in\Sp(2N,\R),~c\in \R^{2N},$$ which acts on the
extended phase vector $\left(\begin{array}{c}
  z \\
  1 \\
\end{array}\right)$.

Symplectic operations executed by CV quantum computers are
particularly important because they correspond to information
processing tasks for which these computers are expected to
outperform their discrete counterparts. In the context of quantum
optics, which is the basis of most proposed schemes for CV quantum
computation, they require only linear optics and squeezing; as
such, these gates may be fairly straightforward to implement
\cite{BraunPati2003}. Theoretically, the generalized
Gottesman-Knill (GK) theorem provides a valuable tool for
assessing the classical complexity of a given continuous quantum
information process \cite{BarSan2002}. It states that any quantum
algorithm that initiates in the computational basis and employs
only the restricted class of affine symplectic gates, along with
projective measurements in the computational basis, can be
efficiently simulated on a classical computer. This computational
model is often described within the stabilizer formalism
\cite{NieChu2000,BarSan2002} as Clifford group computation. To
achieve universal continuous quantum computation, it is necessary
to introduce additional operations (corresponding to elements of
the nonlinear symplectic diffeomorphism group), afforded in the
quantum optics laboratory by the ability to count photons.

Quantum optical control components can include linear elements (such
as beam splitters, mirrors, and half-wave plates), nonlinear
elements (such as squeezers, parametric amplifiers and
down-converters) or a combination thereof. The symplectic
transformations that can be performed by linear optics consist of
the inhomogeneous displacement transformation and the maximal
compact subgroup $\OSp(2N,\R)$ of orthogonal symplectic matrices,
which preserve the total photon number
$$\hat n=\sum_{i=1}^N \hat a_i^\dag \hat a_i.$$
This subgroup is isomorphic to the unitary group $\U(N)$ via the correspondence:
\begin{equation}\label{U-S map}
X-iY\in \U(N)~\rightarrow ~S=\left(%
\begin{array}{cc}
  X& Y \\
  -Y & X \\
\end{array}%
\right).\end{equation} Squeezing operators (also referred to as
active transformations) cannot be implemented using linear optics;
they fall into the noncompact portion of $\Sp(2N,\R)$, and
correspond to photon non-conserving transformations. These operators
rescale canonical operators by a real number $\lmd$ along one axis
in the quadrature plane, and by $\lmd^{-1}$ along the conjugate
axis:
$$\hat q \rightarrow \lmd \hat q,\quad \hat p \rightarrow \lmd^{-1}\hat
p.$$

There exists a set of universal symplectic gates whose combinations
may realize arbitrary Clifford gates in $\ISp(2N,\R)$. A well-known
choice of the universal gate set consists of the Pauli operators,
the Fourier gate, phase gate, and the SUM gate. The Pauli operators
perform the phase displacements
$$X(q_i)=\exp(i\hbar q_i\hat p_i), \quad Z(p_i)=\exp(i\hbar p_i\hat
q_i),$$whose symplectic representations are
\begin{equation}
 X(q) =\left(%
\begin{array}{ccc}
  1 & 0 & q \\
  0 & 1 & 0 \\
  0 & 0 & 1 \\
\end{array}%
\right),
 \quad
 Z(p) =\left(%
\begin{array}{ccc}
  1 & 0 & 0 \\
  0 & 1 & -p \\
  0 & 0 & 1 \\
\end{array}%
\right) \quad \in \ISp(2N,\R).
\end{equation}

The one-qunit Fourier transform is the CV analog of the discrete
Hadamard gate:
$$F = \exp\left\{\frac{i}{\hbar}\frac{\pi}{4}({\hat q}^2+{\hat p}^2)\right\}:~~\left(%
\begin{array}{c}
  \hat q \\
  \hat p \\
\end{array}%
\right) \rightarrow \left(%
\begin{array}{c}
  \hat p \\
  -\hat q \\
\end{array}%
\right).$$ This action can be represented by a $3\times 3$ affine
symplectic matrix
\begin{equation}
 F =\left(%
\begin{array}{ccc}
  0 & 1 & 0 \\
 -1 & 0 & 0 \\
  0 & 0 & 1 \\
\end{array}%
\right)\in \ISp(2N,\R).
\end{equation}
The phase gate is the analog of the discrete variable phase gate
 $$P(\eta) = \exp\left(\frac{i}{2\hbar}\eta{\hat q}^2\right):~~\left(%
\begin{array}{c}
  \hat q \\
  \hat p \\
\end{array}%
\right) \rightarrow \left(%
\begin{array}{c}
  \hat q \\
\hat p - \eta \hat q \\
\end{array}%
\right).$$ Unlike the other gates, $P$ is a function of a real
parameter, and can be represented by the matrix
\begin{equation}
 P(\eta) =\left(%
\begin{array}{ccc}
  1 & 0 & 0 \\
 -\eta & 1 & 0 \\
  0 & 0 & 1 \\
\end{array}%
\right) \in \ISp(2N,\R).
\end{equation}

Finally, the SUM gate acts on a two-qunit system where qunit $1$
is said to be the control and qunit $2$ is said to be the target
\cite{BarSan2002, GotKit2001}, and it carries out the following
transformations on the canonical observable operators:
$$\SUM:\quad\hat q_1 \rightarrow \hat q_1,\quad\hat q_2 \rightarrow \hat q_1 + \hat q_2,\quad\hat p_1 \rightarrow \hat p_1 - p_2,\quad\hat p_2 \rightarrow \hat p_2.$$
This is the continuous-variable analog of the discrete CNOT gate
and its unitary representation is
$$\SUM = \exp\left(-\frac{i}{\hbar}\hat q_1\hat p_2\right).$$
Therefore, the associated symplectic representation is
\begin{equation}\label{sum}
 \SUM =\left(%
 \begin{array}{ccccc}
  1 & 0 & 0 & 0 & 0\\
  1 & 1 & 0 & 0 & 0\\
  0 & 0 & 1 & -1 & 0\\
  0 & 0 & 0 & 1 & 0\\
  0 & 0 & 0 & 0 & 1\\
\end{array}%
\right) \in \ISp(2N,\R).
\end{equation}

\section{Control Landscape Topology for Symplectic Gate Fidelity}
In this work, we are concerned with the optimal control problem of
identifying the time-dependent functional form of the Hamiltonian
that maximizes the fidelity of a symplectic (CV) gate at a fixed
time $t_f$, with a particular focus on the convergence of search
algorithms for this problem.  Prior work \cite{Braunstein2005} has
begun to examine the question of constructing minimal time quantum
circuits for a given symplectic gate from restricted control
Hamiltonians. By contrast, we are interested in the general
problem of gate control for arbitrary quadratic Hamiltonians and
arbitrary final times. This problem must be solved through
computational search over the space of time-dependent quadratic
coupling Hamiltonians that minimize the distance to the target
gate. In continuous quantum computation, we are primarily
interested in the set of gates that can be efficiently simulated
classically, i.e., those that form the inhomogeneous symplectic
group $\ISp(2N,\R)$. Since the (quantum) symplectic gate $U$ is a
faithful unitary representation of a symplectic matrix $S$, it is
reasonable and convenient to define the gate fidelity analogously
to that for discrete gates as
\begin{equation}\label{control landscape}
\J[C(\cdot),t_f]=\tr(S-W)^T(S-W)+(s-w)^T(s-w), \quad S_s\in
\ISp(2N,\R),
\end{equation}where $W_w$ (including a homogeneous transformation $W$
and a $w$-translation) is the finite-dimensional representation of
the target quantum propagator, and $S_s$ is the representation of
the system propagator $U(t)$ as an implicit function of
$C(\cdot)$. Since the control of the inhomogeneous part via linear
couplings has been extensively studied in the literature
\cite{Belavkin1983,BeuRab1990}, we are primarily concerned with
the control of the homogeneous part $S$ and will always omit the
inhomogeneous part (in practice, by switching off the
corresponding linear interactions).

The unitary propagator is an implicit function of the control
field $C(t)$ through the controlled \sd equation
\begin{equation}\label{controlled-sd}
\frac{\partial {U}(t)}{\partial t}=-
\frac{i}{\hbar}~\left[\hat{H}_0(t)+\sum_{i=1}^mC_i(t)\hat
H_{i}\right]{U}(t),
\end{equation}
where $\hat H_0$ is the (quadratic) internal Hamiltonian and $\hat
H_i$ is the interaction Hamiltonian steered by a control field
$C_i(t)$ to couple the internal degrees of freedoms. Hence, the
optimization problem is formally defined on the space of
time-dependent control fields subject to the dynamical constraint
of the \sd equation (\ref{controlled-sd}), which is equivalent to
the following dynamical constraint on $S$:
\begin{equation}\label{controlled-sd-2}
\frac{\dd {S}(t)}{\dd t}=J\left[{H}_0(t)+\sum_{i=1}^mC_i(t)
H_i\right]{S}(t).
\end{equation}Within the
framework of quantum optics, a compact symplectic gate corresponds
to a control implemented via linear optics, and a noncompact gate
corresponds to a control involving implementations of squeezing.

As described in prior work \cite{WuRaj2006,RabMik2005}, any search
for the optimal control fields minimizing such a cost functional
for a given physical system will traverse a so-called control
landscape, defined as the map between admissible control fields
$\{C_i(t)\}$ and the associated values of $\J$. In order to
facilitate our comparison of the optimization efficiencies of
discrete and CV quantum gates, it is useful to acquire the
critical topology for the latter problem, i.e., the distribution
of all possible critical points, including suboptima that may
serve as undesirable attractors for the search trajectory. Since
the cost function is a complete function of the propagator $S$, it
would be convenient to investigate the landscape topology on
$\Sp(2N,\R)$. According to \cite{WuPech2006,WuRab2007}, the
optimality status (i.e., minimum, maximum or saddle point) of a
critical control field $\{C_i(t)\}$ is equivalent to that of the
resulting gate $S$ on $\Sp(2N,\R)$, provided that the map from
$\{C_i(t)\}$ to $S$ is locally surjective at $\{C_i(t)\}$ on the
set of realizable gates at the time $t_f$. Such controls are
conventionally called regular extremal controls. Vanishing of the
gradient $\nabla\J[C(\cdot)]$ may also be caused by singularities
of the map from $\{C_i(t)\}$ to $S$. These critical points, which
are independent of the choice of the cost function, are called
singular extremal controls \cite{BonChy2003}. For simplicity, only
regular extremals are considered, and the singular extremals will
be studied in the future. Actually, in most cases of this paper
the local optima are not singular \cite{BonChy2003}.

In addition, we assume that the system is fully controllable at
the final time $t_f$, i.e., any Clifford group element can be
achieved by some cleverly designed control functions $\{C_i(t)\}$,
so that the system is capable of achieving arbitrary Clifford
group computations. The critical topology problem can then be
reduced from the (infinite-dimensional) domain of  control fields
onto the symplectic group. Generally, a fundamental requirement
for the controllability of systems evolving on Lie
groups\cite{Suss1972} is the rank condition, i.e., the condition
that the Lie algebra spanned by $H_0,H_1,\cdots,H_m$, and their
commutators such as $[H_0,H_i]$, $[H_0,[H_0,H_i]]$,
$[H_i,[H_0,H_j]]$, etc., equals the Lie algebra of the Lie group.
This condition has been proven to be sufficient for the unitary
propagators of discrete-state quantum systems \cite{Suss1972}, and
for these systems there exists a positive time $T_0$ such that
arbitrary gate can be achieved exactly at any positive time larger
than $T_0$. For systems on noncompact symplectic groups, the rank
condition is only sufficient when $H_0$ is compact, and there is
no guarantee of exact-time controllability, i.e., particular gates
may only be reachable after an extremely long time. The
controllability usually fails when $H_0$ is noncompact
\cite{Suss1972,Jurdjevic1997}. However, multiple control fields
may greatly enhance the controllability, such that any gate can be
achieved at arbitrary positive times (rendering the system
strongly controllable), if their corresponding control
Hamiltonians span the whole Lie algebra of the symplectic group.
Unfortunately, many physical models possessing multiple control
fields for CV quantum computing are still not strongly
controllable. However, as discussed in Section IV, it is possible
to design systems with proper coupling Hamiltonians that guarantee
the strong controllability of the system. In what follows, we
assume that a final time $t_f$ has been found at which the system
is controllable, and derive the corresponding control landscape
topology analytically.

It is instructive to review the case of discrete variable quantum
gate control. In this case, a typical measure of fidelity is the
Frobenius matrix norm on $\U(N)$, defining the following objective
function:
\begin{equation}\label{ulandscape} \J(U)=\|W-U\|^2_F=\tr(W-U)^\dag
(W-U)=2N-2Re\tr(W^\dag U).
\end{equation} where $W\in \U(N)$ is the target unitary transformation.
Extensive numerical studies of gate control optimization using
this measure have been reported \cite{PalKos2002}; local iterative
algorithms have generally succeeded in locating the global optima
(albeit at a somewhat higher expense than the optimization of
quantum observables). Application of the tools of differential
geometry on the Riemannian manifold $\U(N)$ \cite{HelMor1994} has
enabled identification of the critical manifolds of this objective
function \cite{Mike2006b}. The essential findings of this work
were 1) no local traps exist in the control landscapes for
discrete quantum gate implementation, consistent with the routine
success of the associated OCT calculations and 2) the critical
topology of these landscapes is universal, i.e., independent of
the target gate $W$ as well as the Hamiltonian. The correspondence
of critical points on the unitary propagator and control field
domains was also the subject of a separate recent investigation,
where the critical topology on $C(t)$ was derived explicitly under
the electric dipole approximation, and found to be essentially
identical to that on the dynamical group $\U(N)$
\cite{HoJason2006,RabMik2005}.

\subsection{Critical
Landscape Topology over the Symplectic Group}  The optimization
problem (\ref{control landscape}) can be decomposed into two
separate problems on the groups $\Sp(2N,\R)$ and $\R^{2N}$,
respectively, where the latter has a unique optimum $s=w$. Thus,
we only the critical topology on $\Sp(2N,\R)$ of the function
\begin{equation}\label{symplectic landscape}
\J(S)=\tr(S-W)^T(S-W), \quad S\in \Sp(2N,\R).
\end{equation}The method to solve the critical points is to perturb the cost function in an arbitrary open subset
and determine when the variation always vanishes irrespective to
the perturbation, which leads to the following condition
\cite{WuRaj2006}:
\begin{equation}\label{b} (S^TS-W^TS)J=J(S^TS-S^TW).
\end{equation}
In contrast to the case of the unitary group, the critical
topology becomes more complex as will be shown below, because
(\ref{symplectic landscape}) does not possess a favorable linear
form as shown in (\ref{ulandscape}). Nevertheless, this equation
is still analytically solvable owing to its highly symmetric form.
Here we only present the conclusions, and the details can be found
in a separate work of the authors \cite{WuRaj2006}. Let $W=UEV$ be
the singular value decomposition of $W$, where $U,V\in\OSp(2N,\R)$
and the diagonal matrix $E$ contains the singular values of $W$.
Suppose that there are $2n_0$ singular values $e_0=1$ (if
existing) and the remaining singular values $1<e_1<\cdots<e_s$
have degeneracies $n_1,\cdots,n_s$. In the context of quantum
optics, the pre-transformation
 $V$ and the post-transformation $U$
decompose the target symplectic transformation into operations on
decoupled modes, which includes
 $n_0$ linear operations and the remainder are squeezing operations with
 squeezing ratios $e_1,\cdots,e_s$ on the $\hat p$-components.
In the following discussions, we always assume that $U=V=I_{2N}$
so that the physics can be discussed in a simple canonical
coordinate system where all the modes are decoupled.

The critical submanifolds can be expressed as \cite{WuRaj2006}:
\begin{equation}\label{critical points}
S^*=R^TDR,\quad R\in \G(E),\end{equation} where the stabilizer
$\G(E)$ of $E$ in $\OSp(2N,\R)$ is defined as
$$\G(E)=\{R\in \OSp(2N,\R)|R^TER=E\}=\OSp(2n_0)\times O(n_1)\times \cdots O(n_s).$$
The characteristic matrix $D$ consists of different operations on
the separate modes represented by the following diagonal blocks
(their inverses appear pairwisely symmetrically in $D$ and $E$
according to the reciprocal property of eigenvalues of symplectic
matrices):

  {(1)} Type-I operations $D'_\alpha=e_\alpha I_{m'_\alpha}$ corresponding to a sub-block $E'_\alpha=  e_\alpha I_{m'_\alpha}$ in $E$, which are
  identical with those operations in the target transformation $W$;

  {(2)} Type-II operations $D''_\beta= -e_\beta^{1/3} I_{m''_\beta}$ corresponding to a sub-block $E''_\beta=e_\beta^{-1} I_{m''_\beta}$ in $E$,
  which reverse the directions of the quadrature
  vector of the corresponding modes and is followed by a squeezing operation with ratio $e_\beta^{1/3}$ on the $\hat
  q$-components.

  {(3)} Type-III operations $$D_{\gamma\delta} =\sqrt{\frac{e_\delta}{e_\gamma}}\left(%
\begin{array}{rr}
\cos x_{\gamma\delta}I_{m_{\gamma\delta}}    & \pm\sin x_{\gamma\delta}I_{m_{\gamma\delta}} \\
\pm\sin x_{\gamma\delta}I_{m_{\gamma\delta}}   & -\cos x_{\gamma\delta}I_{m_{\gamma\delta}} \\
\end{array}%
\right),$$where $x_{\delta\gamma}=\arccos\frac{{e_\gamma}/{
e_\delta}-{e_\delta}/{e_\gamma}}{(e_\delta e_\gamma)^{\frac{1}{2}}
-( e_\delta e_\gamma)^{-\frac{1}{2}}}$, corresponding to a sub-block $$E_{\gamma\delta}=\left(%
\begin{array}{rr}
e_{\gamma}I_{m_{\gamma\delta}}    &  \\
   & e_{\delta}^{-1}I_{m_{\gamma\delta}} \\
\end{array}%
\right),~~~e_\delta^{1/3}\leq e_\gamma \leq e_\delta,$$which
rotates two decoupled $\gamma$-th and $\delta$-th modes with an
angle $x_{\delta\gamma}$ followed by a uniform squeezing operation
on both modes.

(4) The sub-block for the particular singular value $e_0=1$ is
$$D_0=\left(%
\begin{array}{cc}
  I_{m_0} &    \\
   & -I_{n_0-m_0}  \\
\end{array}%
\right),$$which leaves $m_0$ modes of linear operations invariant
and reverses the direction of quadrature vectors of the other
$n_0-m_0$ modes in the phase space.

Moreover, the critical transformations allows for arbitrary linear
optical transformations in the stabilizer $\G(E)$ as shown in
(\ref{critical points}). Each combination of the above operations
forms a critical submanifold, and the indices
$\{m_0,m'_\alpha,m''_\beta,m_{\gamma\delta}\}$ label each
individual critical submanifold. All admissible combinations can
be enumerated to count the number of the critical submanifolds.

Having identified the complete set of critical points, it is
important to determine their optimality status (i.e, maxima,
minima or saddle points) through analysis of the eigenvalue
structure of the Hessian quadratic form, so as to understand their
influence on the search effort for optimal controls. {Enumeration
of the positive, negative and zero eigenvalues of the Hessian at a
critical point provides information about the number of the
upward, downward and flat orientations of the principal axis
directions of the landscape, respectively. The counting results
are given in \cite{WuRaj2006}; they show that the local minimum
$S^*=W$ is unique among all the critical solutions, and the
remaining critical points are all saddle points. Thus, no false
traps are present to impede the search of optimal controls.}

\subsection{Landscape critical topology on $\OSp(2N,\R)$}

Carrying out optimal control field searches using only linear
quantum optics corresponds to searching over only the compact
subgroup $\OSp(2N,\R).$ Such a strategy may be desirable in the case
of compact symplectic gates, such as $F, X$ or $Z$ above. {\it A
priori}, it is not obvious when it is desirable to search over only
this subgroup versus the whole group using a combination of linear
optics and squeezing. Since such decisions would be facilitated by
knowledge of the landscape topology over $\OSp(2N,\R)$, we also
carry out a critical point analysis over this domain.

Firstly, the landscape function is greatly simplified on
$\OSp(2N,\R)$:
\begin{equation}\label{}
\J(S)=\tr(S-W)^T(S-W)=4N-2\tr(W^TS), \quad S\in \OSp(2N,\R),
\end{equation}
where $W$ is the target compact symplectic gate. In fact, by
(\ref{U-S map}), the control landscape over $\OSp(2N,\R)$ can be
mapped to an equivalent control landscape in the form of
(\ref{ulandscape}). Hence, it is obvious \cite{WuRaj2006} that the
critical topology is identical with that of the unitary
transformation landscape \cite{RabMik2005}. Specifically, the
critical manifolds are $\OSp(2N,\R)$ orbits \cite{WuRaj2006},
i.e., $S^*=WO^TD_mO$, where $O\in\OSp(2N,\R)$ and
$$D_m=\left(%
\begin{array}{cccc}
  -I_m & & &  \\
   &  I_{N-m} & & \\
   &   &  -I_m &\\
   &   &   &  I_{N-m} \\
\end{array}%
\right),\quad m=0,1,\cdots, N.
$$
Hence, there are a total of $N+1$ critical solutions, with values
of the cost functional $\J=0,8,16,...,8N$. The minimum and maximum
values of $\J$ correspond to $S=W$ and $S=-W$, respectively, and
the rest critical points are saddle.

We note the important fact that whereas the dimension of a unitary
transformation representing a $N$-qubit discrete quantum gate
scales exponentially as $2^{N}$ \cite{NieChu2000}, for continuous
systems, the dimension of the symplectic transformation required
to implement the equivalent continuous quantum gate within the
Clifford group scales linearly as $2N$, where $N$ is the number of
qunits. This is understandable because the computational
capabilities of discrete quantum computers extend beyond those of
Clifford-gate CV quantum computers.

\section{OPTIMAL CONTROL OF CONTINUOUS VARIABLE GATES}
Algorithms for quantum optimal control have been extensively
developed for the control of discrete (finite-dimensional) quantum
systems, or continuous quantum systems that can be treated as
finite-dimensional to a reasonable approximation. Broadly
speaking, two distinct types of discrete variable quantum control
problems have been considered from an algorithmic point of view:
1) control of quantum observables; 2) control of dynamical
propagators (gates). It has been noted that the latter is
generally computationally more expensive, in part because the
solution set is relatively small to locate. One of our primary
aims in this work is to characterize the expense of solving 3)
optimal control problems for CV quantum dynamical propagators
under various constraints and to compare this expense to that of
2). For similar reasons, one would expect problem 3) to be
inherently more difficult than the optimization of CV observables.

The analysis above shows that despite the noncompactness of the
symplectic group of CV propagators, it is possible to choose a
scalar fidelity function with a fairly simple critical topology. In
this section, we describe the physical models employed for CV
quantum computations and the numerical algorithms used to search for
optimal controls.

\subsection{Physical control systems for CV quantum information
processing}

{Over the past few years, several physical models have been
suggested for the implementation of symplectic gates. Early
proposals focused on coupled pairs of conjugate continuous
variables describing quadrature modes of the electromagnetic field
\cite{GotKit2001,LloBra1998,Fiurasek2003}. In such traditional
quantum optics models, it is typically possible to apply only one
control at any given time. Although it is often possible to
implement CV gates via one control with reasonable fidelity if the
final time $t_f$ is chosen judiciously, the quality of control
will be downgraded. In the present work, we do not limit ourselves
to these restricted control Hamiltonians from quantum optics,
since our goal is to explore the generic properties of the CV gate
optimal control problem.}

{More recently, CV models displaying greater flexibility have been
proposed that may be more suitable for optimal control of quantum
gates. In particular, these control systems allow for the
simultaneous application of two or more independent controls.} A
simple model raised in \cite{Kraus2003,Fiurasek2003} considered
the off-resonant interaction of light with a collective spin
described by the effective Hamiltonian
$$H_{0}=\kappa x_1p_2$$which represents a strongly coherent light beam
polarized along the $x$-axis that propagates along the $z$-axis
through the atomic ensemble. In addition, it is assumed that
arbitrarily fast local phase shifts are implementable by
single-model control Hamiltonians:
$$H_{1}=x_1^2+p_1^2,~~~~H_{2}=x_2^2+p_2^2.$$
This system satisfies the rank condition, but is not ensured to be
fully controllable because $H_0$ is noncompact. In
\cite{Fiurasek2003}, control pulses are restricted to be
instantaneous and exerted in certain sequences to simplify the
analysis and the experimental realization. However, here we will
assume that the control pulses can be arbitrarily shaped, so that
a greater degree of precision is possible in tailoring the control
Hamiltonian to match theoretical predictions.

Recently, atomic ensembles, particularly ensembles of trapped
ions, have emerged as a promising medium for CV QIP
\cite{Kim2005,Huang2006,Parkins2006}, because the trapped ions are
thermally isolated from their environment, minimizing decoherence
effects. {These systems may also offer a degree of flexibility
suitable for gate synthesis via OCT.} For CV gates, quantum
information is stored in the vibrational modes of the trapped
ions. To couple (entangle) these vibrational modes, several
studies have examined the interactions of vibrational states of
trapped ions with some quantized fields inside an optical
cavity\cite{Kim2005,Huang2006}, through which it is possible to
indirectly tune the coupling between vibrational modes.

For concreteness, consider a model wherein two trapped ions with
internal electronic levels are coupled to external lasers. They
are also coupled to a cavity mode with frequency $\omega_c$,
described by annihilation and creation operators $a$ and
$a^{\dag}$, respectively, where the harmonic frequency of each
trap is $2\nu$. We assume that the cavity is oriented along the
$x$ axis and the laser is incident along the $y$ or $z$ axis. Then
in a frame that is rotating with frequency $\omega_c$, the
interaction Hamiltonian coupling the vibrational modes of the ions
with the cavity and laser fields can be written (omitting the
electronic states):
$$H = 2\nu( b_1^{\dag}b_1+ b_2^{\dag}b_2) +V,$$ where $b_j$ and $b_j^{\dag}$ ($j=1,2$) are
the annihilation and creation operators of the vibrational modes.
The interaction Hamiltonian $V$ is a function of the coupling
constants between ions and lasers and single-photon coupling
strengths. Under reasonable assumptions regarding the size of the
traps compared to the laser wavelength, and with proper detuning
of the lasers from the cavity mode, it can be shown
\cite{Huang2006} that the interaction Hamiltonian for the first
ion can be approximated as
$$V\approx r_{11}(a^\dagger b_1+ab_1^\dagger)+
r_{21}(a^\dagger b_2^\dagger+ab_2).$$ where the $r_{mn}$ are
functions of the frequencies and coupling constants. This system
actually involves three harmonic oscillators, where ion-cavity
interaction Hamiltonians produce indirect couplings between the
vibrational modes of the two ions. By modulating the frequencies
of these lasers through time, we can achieve time-dependent
control Hamiltonians necessary for the implementation of CV gates
with optimal fidelity. The controls in this model induce nonlocal
interactions between qunits. The associated control system does
not satisfy the controllability rank condition, and hence is
uncontrollable.

The above models display features that are representative of
current proposals for CV QIP. As we will see below, their
controllability properties are of particular importance. The
light-collective spin interaction control system (hereafter
referred to as the "photon model") is sufficient for achieving
arbitrary symplectic transformations in experiments, but its
controllability is not strong enough to achieve the target in
arbitrary finite time. On the other hand, since the ion trap model
is not fully controllable, for certain gates, there does not exist
a final time $t_f$ at which the gate can be reached with arbitrary
precision. Nonetheless, for many gates, it is possible that such a
final time exists.

In addition to these two physical models, we also choose a
strongly controllable system below in order to explore the effects
of controllability on the properties of gate optimization. The
following control Hamiltonians were employed for this purpose:
$$\hat H_{1}=\hat a_1^{\dag 2}-\hat a_1^2+\hat a_2^{\dag 2}-\hat a_2^2,
 \quad \hat H_{2}=\hat a_1^\dag \hat a_1-\hat a_2^{\dag}\hat a_{2}+i(\hat a_{1}^{\dag}+\hat a_1)(\hat a_{2}^{\dag}-\hat a_2);$$
the internal Hamiltonian in this case consisted of uncoupled
harmonic oscillators, i.e. $H_0=\hat a_1^\dagger \hat a_1+\hat
a_2^\dagger \hat a_2+1$.

As in the case of discrete variable quantum control, it is
possible to impose additional constraints on the optimization
problem based on the physical implementation of choice, such as
bounds on the control intensity or the time derivative of the
control pulse. A study of the impact of such constraints is
properly the subject of a separate work.

In the following, we make several comparisons to discrete QIP. For
this purpose, we assume the standard physical model of NMR-based
quantum computation \cite{Khaneja2001}. In this model, the
internal Hamiltonian $H_0$ consists of nuclear spins that are
coupled in the absence of the control field. The coupling between
$N$ spins (only up to 2-qubit interactions are considered) is
achieved through standard NMR coupling Hamiltonians of the form:
$$H=\sum_{i=1}^N\omega_i\sigma_{i}^z+\sum_{i,j=1}^N\sum_{\alpha,
\beta=x,y,z} J_{ij}^{\alpha,\beta}\sigma_{i}^\alpha\otimes
\sigma_j^{\beta}+\sum_{i=1}^NC_i(t)\sigma_{i}^x.$$ where
$\sigma_i^{x,y,z}$ are the standard Pauli matrices representing
observables of the $i$-th qubit. The first term splits the energy
levels via a static magnetic field along the $z$-axis, with
$\omega_i$ being the Raman frequencies; the second term represents
the internal couplings between the qubits (e.g., chemical shifts);
the last term, the control Hamiltonian, interacts each qubit states
to a time-variant $x$-axis radio-frequency control field $C_i(t)$.
Because of the tensor product structure of qubit subspaces in these
expressions, the total system dimension scales as $2^N$. One can
verify that this system is controllable, but not strongly
controllable.

\subsection{Numerical implementation}

Several optimization algorithms, including iterative methods such
as the Krotov algorithm \cite{PalKos2002}and tracking methods such
as D-MORPH \cite{RotRab2005}, have been applied in the OCT of
discrete quantum gate implementations. These algorithms can vary
considerably in optimization efficiency, but they are all based on
information pertaining to the first functional derivative of the
objective function with respect to the control field. Since our
primary goal in this work is to compare the properties of discrete
and continuous gate OCT, we adopt gradient algorithms to search
for optimal controls, which, although not the most efficient,
offer the simplest and most direct opportunity for comparison. In
particular, they are ideal for detecting landscape traps. By
comparing the magnitudes of the gradient and Hessian of the
objective function along the search trajectory, we can obtain a
understanding of the factors that govern optimization efficiency
and what can be applied to the design of tailored algorithms in
future work.

The electric field $C_i(s,t)$ was stored as a $p\times q$ matrix,
where $p$ and $q$ are the number of discretization steps of the
algorithmic time parameter $s$ and the dynamical time $t$,
respectively. For each algorithmic step $s_k$, the field was
represented as a $q$-vector for the purpose of computations.
Starting from an initial guess $C_i(s_0,t)$ for the control field,
the equations of the motion were integrated over the interval
$[0,t_f]$ by propagating the \sd equation over each time step
$t_k\rightarrow t_{k+1}$, producing the local propagator
$U(t_{j+1},t_j) = \exp[-iH(s_i,t_j)t_f/(q-1)]$. The method used
for this purpose differed for the discrete quantum and symplectic
gate optimization problems. For discrete quantum systems, the
Hamiltonian matrix was diagonalized (at a cost of $N^3$), followed
by exponentiation of the eigenvalues, and multiplication of the
resulting matrix on the left and right by the matrix of
eigenvectors and its transpose. Alternatively, a fourth-order
Runge-Kutta integrator can be employed for the propagation. The
propagation toolkit, which involves precalculating the matrix
exponential at discrete intervals over a specified range of field
amplitudes, was used to further improve the speed of Hamiltonian
integration for discrete quantum systems. The initial guess
$C(s_0,t)$ was alternatively taken as a random, constant, or sin
pulse field.

The Pade approximation for the exponential function was used to
calculate the local symplectic propagators $S(t_{j+1},t_j) =
\exp[JH(s_i,t_j)t_f/(q-1)]$. Since the matrix $JH(s_i,t_j)$ is not
symmetric, it is not possible to calculate its exponential via
diagonalization and subsequent scalar exponentiation of its
eigenvalues. The type $(p,q)$ Pade approximation for $e^x$ is the
$(p,q)$-degree rational function $P_{pq}(x) \equiv
N_{pq}(x)/D_{pq}(x)$ obtained by solving the algebraic equation
$\sum_{k=0}^{\infty}x^k/k!-N_{pq}(x)/D_{pq}(x)=O(x^{p+q+1})$,
i.e., $P_{pq}(x)$ must match the Taylor series expansion up to
order $p+q$. The primary drawback of the Pade approximation is
that it is only accurate near the origin, so that the
approximation is not valid when $JH(s_i,t_j)$ is too large and
when its eigenvalues are not too widely spread. For the problems
considered in this work, the norm of $JH(s_i,t_j)$ was always
small enough for high accuracy in the approximation. Because of
the noncompactness of the symplectic group, which results in the
functional derivatives of the objective function changing too
rapidly, Runge-Kutta integration was not used. Due to the large
number of iterations generally required for convergence of CV
quantum controls, the speed of the matrix exponentiation algorithm
is particularly important. However, the propagation toolkit was
generally found to be inadequate for speeding up symplectic matrix
propagation; discretization of the control field amplitudes
produced unacceptable errors in the matrix exponential, and the
maximum amplitudes often grew abruptly during optimization.

Minimizations of the fidelity function were typically done using the
Polak-Ribere variant of the conjugate gradient (CG) method. Step
size was varied adaptively based on Brent's method for line
minimizations. In several cases, adaptive step size steepest descent
was employed to analyze the behavior of gradient flow lines.
Steepest descent was found to converge much less efficiently than CG
for the symplectic gates. Its performance for discrete unitary gates
was considerably better. For these algorithms, the gradients of the
objective functions were calculated analytically via the following
expressions:

$$\frac{\delta J(S)}{\delta C_i(t)} = \tr~[H_{i}(t)J(S^T(t_f)W-S^T(t_f)S)~]$$
where $H_{i}(t) = S^T(t)H_{i} S(t),$ for CV gates,
and$$\frac{\delta J(U)}{\delta C_i(t)}
={i}\tr~[H_{i}(t)(U^{\dag}(t_f)W - W^{\dag}U(t_f))~],$$ where
$H_{i}(t) = U^{\dag}(t)H_{i} U(t)$, for discrete unitary gates.
Here, the $H_{i}$'s are the Hamiltonians that couples to the
time-dependent control; for molecular systems, it represents the
dipole moment operator.

Integration of the gradient flow equations on the domain of
symplectic or unitary propagators was carried out using the
fifth-order adaptive step size Runge-Kutta method.

\subsection{Impact of landscape topology on numerical control search}

From the critical topology analysis, it is clear that optimal
control landscapes for symplectic transformations are devoid of
suboptimal local traps, regardless of the structure of the gate
transformation or the Hamiltonian of the system. Additionally, all
 critical points stay in a bounded region in the symplectic
group, i.e., they can affect the optimal search only when it
enters that region.

The compactness and degeneracies of the singular values of $W$
determine the critical topology of the control landscape. When $W$
is compact (e.g., $F$, $X$ and $Z$ or combinations thereof), the
number of critical manifolds in the control landscape scales
linearly as $N+1$, where $N$ is the number of qunits. When $W$ is
noncompact and has fully degenerate singular values (i.e.,
$e_i=e^*\neq 1$), the number of critical submanifolds can be shown
to be quadratic in $N$ \cite{WuRaj2006}:
\begin{equation}\label{enum}
  \n=\left\{\begin{array}{ll}
  (N+2)^2/2, & ~N~~{\rm even}; \\
(N+1)(N+3)/2, &~ N ~~{\rm odd}; \\
\end{array}\right.
\end{equation}
The scaling for nondegenerate gates shoots up when the degeneracy
is broken. For the case that $W$ has fully non-degenerate singular
values, the upper bound for the number of critical submanifolds is
$$\n_1=\sum_{m=1}^{[N/2]}\frac{2^{N-3m}N!}{m!(N-2m)!},$$
which is super-exponential.

\begin{figure}[h]
\centerline{
\includegraphics[width=4.5in,height=2.5in]{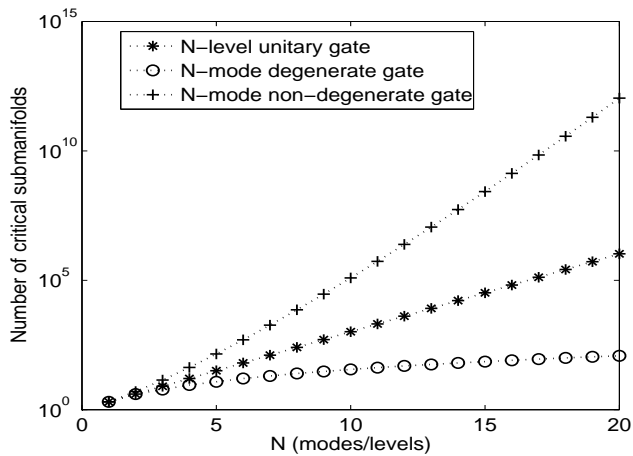}
} \caption{The scaling of the numbers of critical submanifolds for
fully non-degenerate ($N$-qunit) symplectic gates and ($N$-qubit)
unitary gates. }\label{numbers}
\end{figure}

By contrast, it was previously shown that the number and critical
values of all discrete quantum logic gates are independent of the
gate and depend only on the dimension of the system
\cite{RabMik2005}. Fig.\ref{numbers} compares the scalings
calculated above with that of ($N$-qubit) unitary gates. In most
cases, the number of critical submanifolds for symplectic gates
grows much faster than that for unitary gates. However, the number
of critical submanifolds is always finite, and hence the critical
region is bounded for arbitrary target gates $W$, and contained in
the ball centered at $W$ with radius
$$R=\sqrt{\sum_{i=1}^N(e_i^2+e_i^{-2}+3e_i^{2/3}+3e_i^{-2/3}) }\geq 2\sqrt{2N}$$
equal to the distance to the farthest critical points. The volume of
this region is roughly of the order $V\sim R^{2N^2+N}$. Assume that
the attraction is effective in a $r$-ball around each critical
point; then the ratio $\sigma$ of the volume of the attractive
regions to the volume of the region of critical points
$$\sigma\sim\frac{\n\times
r^{2N^2+N}}{R^{2N^2+N}}<\n_1\times\left(\frac{r}{R}\right)^{2N^2+N},$$
can be numerically proven to go rapidly to zero when $N$ approaches
to infinity for arbitrary $r<R$, impling that the probability of a
random initial guess starting close to a saddle manifold should be
so small as to be negligible in practice.

{For a compact target symplectic gate, the critical submanifolds
are completely identical when the search is carried out on the
full symplectic group (corresponding to linear optics plus
squeezing) or its compact subgroup (corresponding to only linear
optics),} the only difference being an increase in the dimension
of the search space. Therefore, the additional directions
accessible through squeezing transformations are not expected to
improve convergence toward the optimal solution. These
observations collectively paint a fairly simple picture of CV gate
landscape topology, which, although more complicated than the
topology of discrete gate landscapes, should not preclude
efficient control optimization.

\begin{figure}[h]
\centerline{
\includegraphics[width=6in,height=2in]{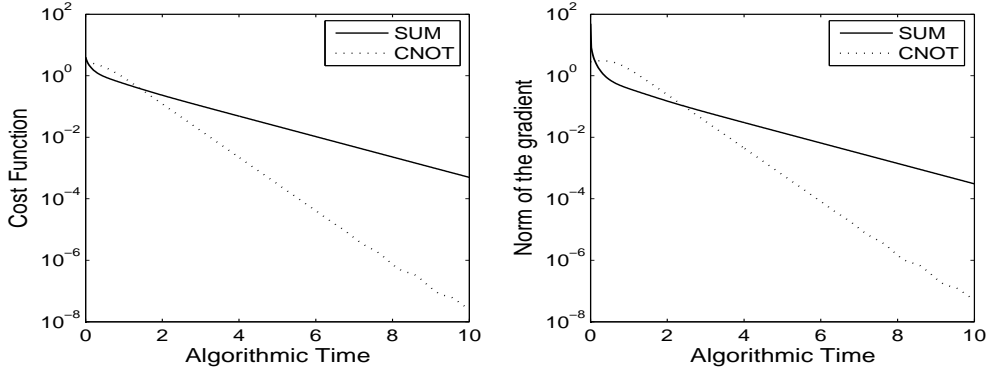}
} \caption{The convergence of the gradient flows for optimal search
of the SUM gate on $\Sp(4,\R)$ and CNOT gate on $\U(4)$.
}\label{kinsumcnot}
\end{figure}

Throughout this paper, we will use gradient algorithms to optimize
the control field. The search process in the kinematic picture can
be represented by the so-called gradient flow on the symplectic
group:
$$\frac{\dd S}{\dd s}=-\nabla \J(S).$$
It is instructive to estimate the convergence speed of the
gradient flow, via linearization of this equation in a
sufficiently small neighborhood of the global optimum $S=W$, which
gives
$$\frac{\dd~ \delta S}{\dd s}=-\h(W)~\delta S,$$
where $\delta S$ is the deviation of $S$ from $W$, which is
proportional to the gradient vector at $S$. The positive definite
matrix $\h(W)$ is the Hessian matrix at $W$. Therefore, the
convergence of $\delta S$ is exponential and its rate is dominated
by the smallest eigenvalue of $\h(W)$, identified as $e_s^{-2}\leq
1$. {By comparison, a similar estimate for gate search on the
discrete unitary group reveals a constant convergence rate of $1$.
Therefore, search for noncompact symplectic gates will display
slower convergence in general, depending on the magnitudes of the
singular values of the target gate. In particular, the convergence
speed for the phase gate (or squeezing gate) decreases with
increasing phase shift (or squeezing ratio).} Fig.\ref{kinsumcnot}
shows the convergence speed of gradient flows for the SUM gate on
the symplectic group, and, for comparison, the CNOT gate on the
unitary group.

\section{Numerical Simulations}

In this section, we numerically solve for optimal control fields for
Clifford group gates and composite CV algorithms, and compare the
optimization effort and control field complexity to those of the
corresponding gates over discrete variables. As such, we aim to
identify distinguishing features of CV gate control that will have
the greatest impact on its computational and experimental
implementation.

\subsection{SUM Gate}
\begin{figure}[h]
\centerline{
\includegraphics[width=6in,height=2in]{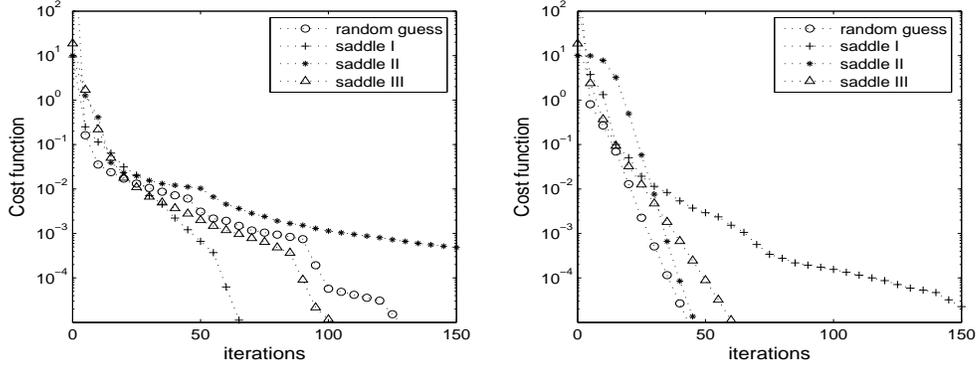}
} \caption{The convergence of dynamical search for the SUM gate
using conjugate gradient algorithms with the photon model (left)
and a strongly controllable system(right).}\label{sumj}
\end{figure}

\begin{figure}[h]
\centerline{
\includegraphics[width=6in,height=4in]{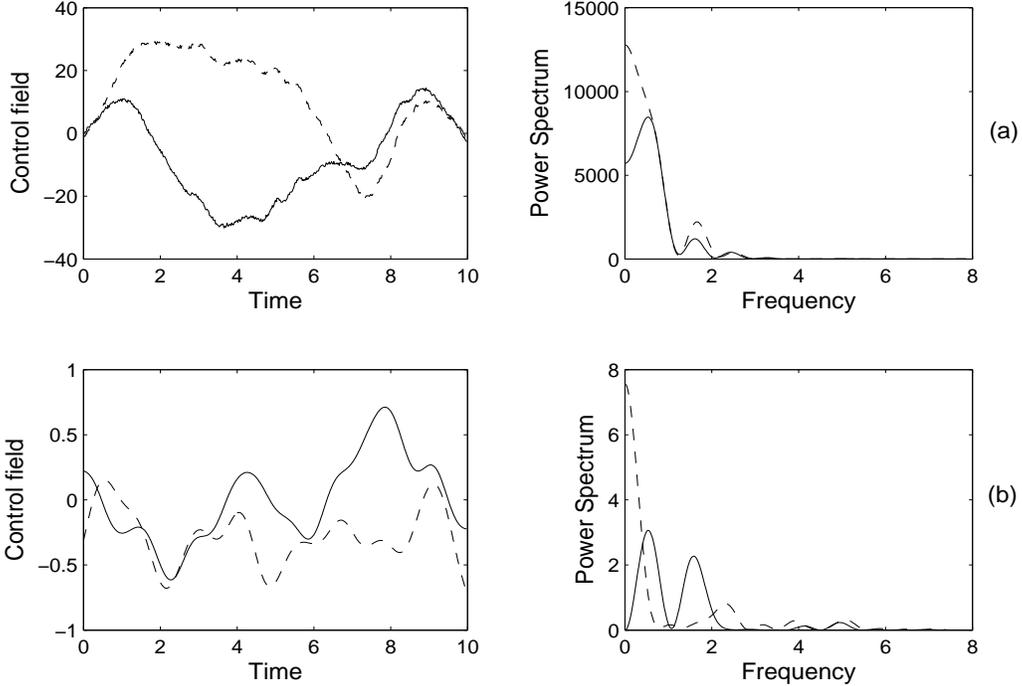}
} \caption{The optimal control fields for CV SUM gate control
after searching from a random initial guess, for (a) the photon
model and (b) a strongly controllable model.}\label{sumf}
\end{figure}

Consider the control problem where $W$ is the SUM operation, whose
matrix form is shown in (\ref{sum}). We simulate the realization
of the SUM gate using both the photon model ($\kappa=1$) {and the
strongly controllable system described above. Fig.\ref{sumj}
compares the effects of strong versus weak controllability on
convergence speed,starting from either a random guess or near
saddle solutions. Although the SUM gate is reachable with both
models at the final time, the weaker controllability of the photon
model compromises convergence speed.} In addition, Fig.\ref{sumf}
shows that the corresponding optimal control fields are more
expensive in that their fluences are much greater.

The singular values of $\SUM$ are
$E=diag\{1.618,1.618,0.618,0.618\}$. The analysis above predicts
that there should be 4 critical submanifolds for a degenerate
2-qunit gate. The first one $D_1=diag\{1.618,1.618,0.618,0.618\}$
contains one type-I block, whose corresponding critical
submanifold is the isolated global minimum $S^*_1={\rm SUM}$. The
second, $D_2=diag\{-0.852,-0.852,-1.174,-1.174\}$, contains one
type-II block, whose corresponding critical submanifold is an
isolated saddle point. The third, $D_3=diag\{1.618,-0.852,-1.174,
0.618\}$ contains one type-I and one type-II block. The last
saddle contains a type-III block, which actually contains two
isolated points. In summary, there are a total of 5 critical
submanifolds, including 4 isolated points and 1 one-dimensional
manifold. The Hessian analysis is summarized in Table I.
\begin{center}
\begin{table}[h]
\begin{tabular}{c|c|c|c|c|c}
\hline
No. & Critical value  & ~~$D_0$~~ & ~~$D_+$~~ & ~~$D_-$~~ & ~type~ \\
\hline
1&        0  &   0 &  14 &        0 & minimum \\
2&   18.623  &   0 &  10 &   4 & saddle \\
3&  9.311    &   1 &  12 &   1 & saddle \\
4($\pm$)&  10       &   0 &  11 &   3 & saddle \\
\hline                                   
\end{tabular}
\caption{Critical topology for the SUM gate.}
\end{table}
\end{center}

As discussed above, the saddles will rarely be encountered during
the progress of most optimization trajectories. This is also
supported from the simulation result (Fig.\ref{sumj}), the saddle
manifolds appear to have a slightly adverse effect on optimization
efficiency for the strongly controllable system, and almost have
no influence on the convergence of the photon model.

\begin{figure}[h]
\centerline{
\includegraphics[width=6in,height=2in]{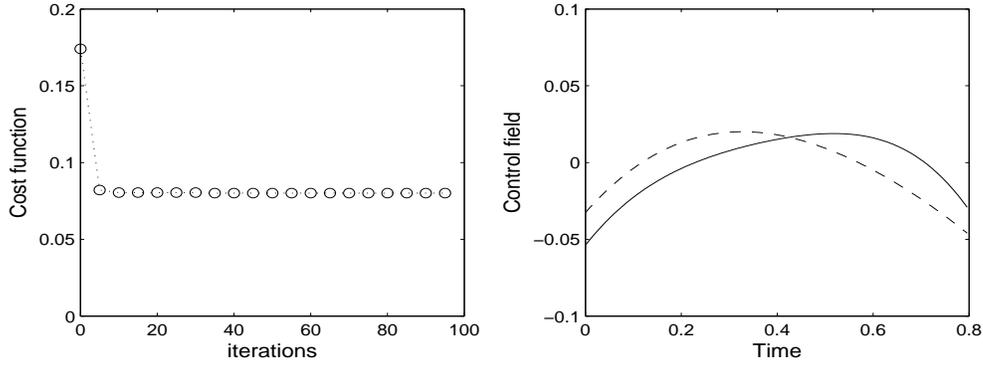}
} \caption{Optimal control search for the SUM gate using conjugate
gradient algorithms with the photon model at a final time smaller
than $\kappa^{-1}$.}\label{sum0.8w}
\end{figure}

\begin{figure}[h]
\centerline{
\includegraphics[width=6in,height=2in]{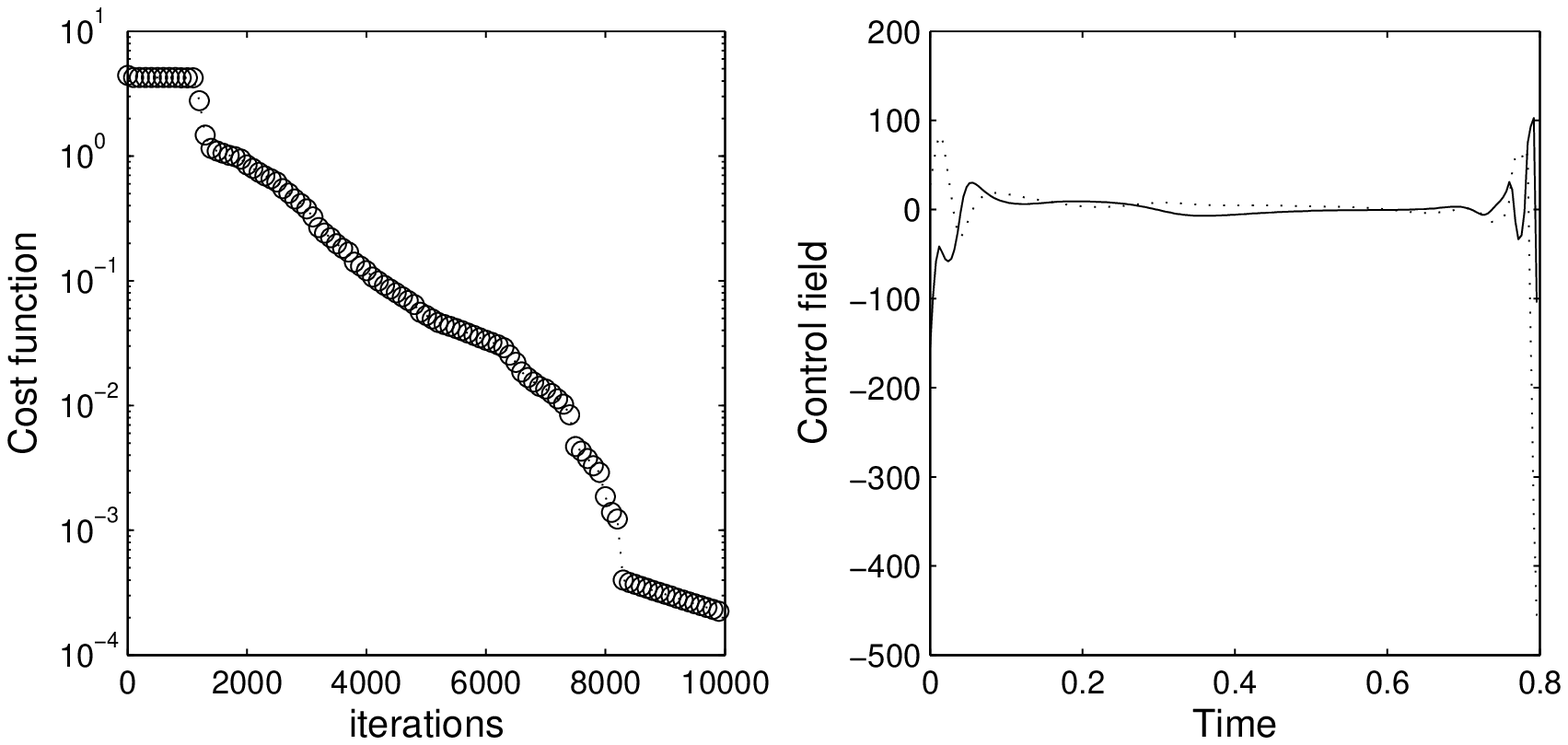}
} \caption{Dynamical search for the SUM gate using conjugate
gradient algorithms with the strongly controllable system model at a
small final time.}\label{sum0.8s}
\end{figure}

Note that the free Hamiltonian for the photon model happens to be
proportional to the matrix logarithm of the SUM gate; thus, using
this model, SUM can be achieved merely via free evolution in
$\kappa^{-1}$ units of time. Simulations show that the SUM gate is
always realizable in any time longer than $\kappa^{-1}$ (e.g.
Fig.\ref{sumf}). However, it is interesting to see if the local
controls permit the gate to be achieved in a shorter time.
Fig.\ref{sum0.8w} shows an example employing a final time less
than $\kappa^{-1}$. As can be seen, the control search does not
converge. By contrast, for the strongly controllable system,
optimal control fields exist even for very small $t_f$, although
the expense increases and the shape of control fields tends to
become more singular (Fig.\ref{sum0.8s}).

An important feature not shown in the figures is that the optimal
search suffers from serious numerical instabilities when it starts
far away from $W$ because the system dynamics involves
exponentially increasing components. This can easily occur when
there is no {\it a priori} knowledge of an appropriate choice for
the control fields. By contrast, because of the compactness of the
dynamical group, the control optimization for discrete quantum
systems never encounters this problem.

\subsection{SWAP gate}
Optimal control landscapes over $\OSp(2N,\R)$ and $\U(N)$ have
identical topology and geometry, because these dynamical groups
are isomorphic. This section simulates the optimal search of a
(compact) transform that swaps the states of the two qunits as
follows
$$W=\left(%
\begin{array}{cccc}
 0  & 1 & 0 & 0 \\
 1   & 0  & 0 & 0 \\
 0  & 0 & 0 & 1 \\
 0  &  0 & 1 & 0 \\
\end{array}\right).$$

\begin{figure}[h]
\centerline{
\includegraphics[width=3.5in,height=2.5in]{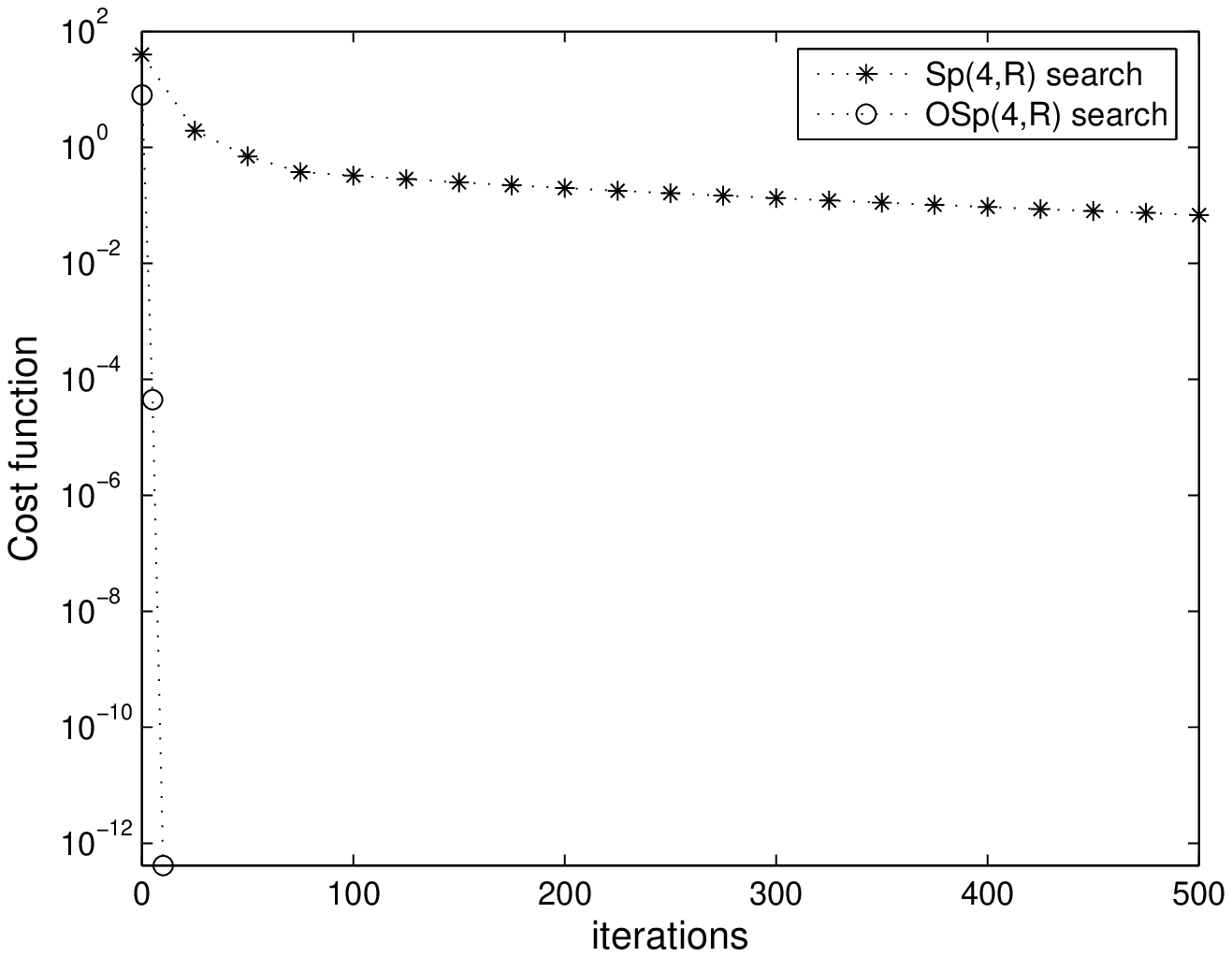}
} \caption{The convergence of optimal search for the SWAP gate
with linear and squeezing couplings ($\Sp(4,\R)$) and linear
couplings ($\OSp(4,\R)$).}\label{swapj}
\end{figure}

\begin{figure}[h]
\centerline{
\includegraphics[width=6in,height=4in]{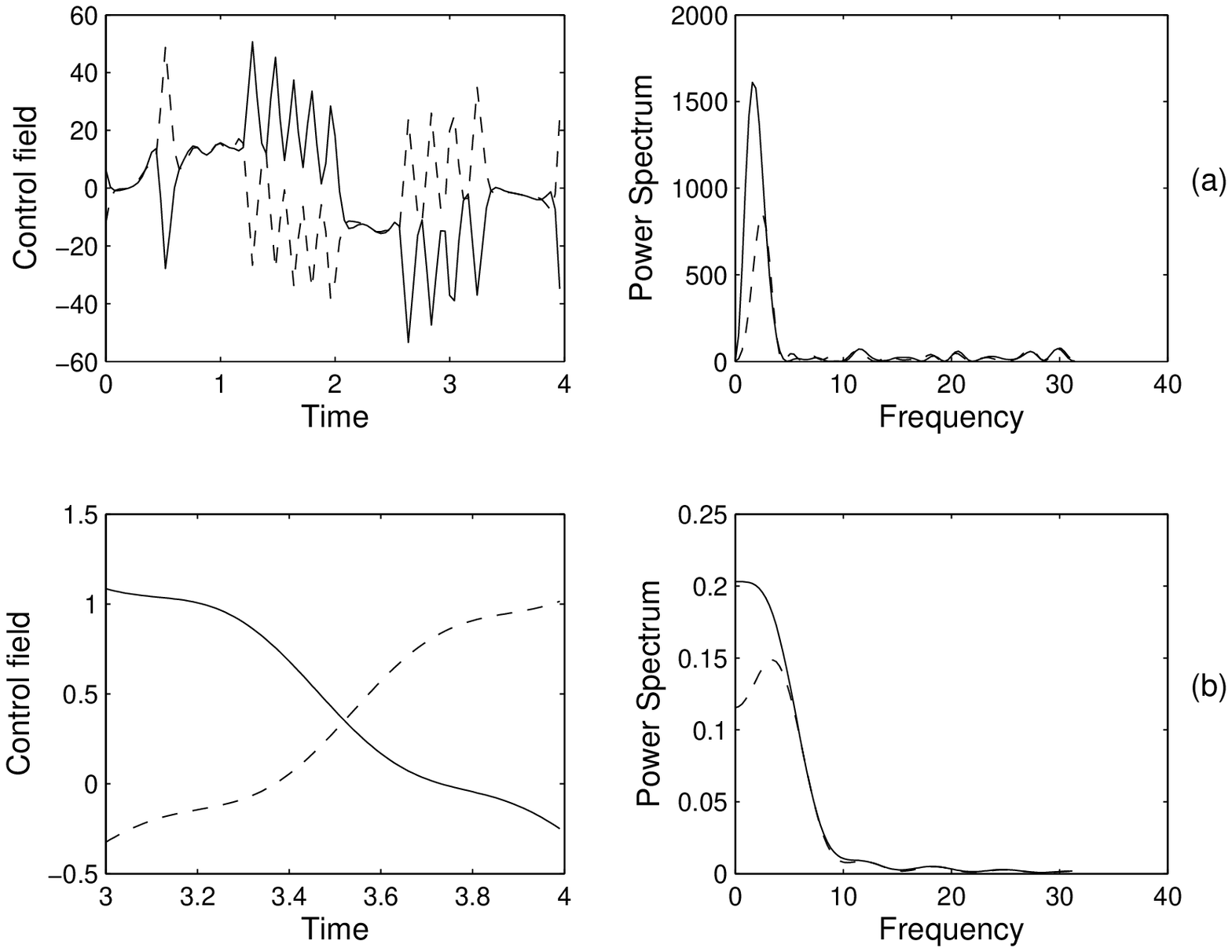}
} \caption{The control fields for the swap gate with (a) linear
and squeezing couplings ($\Sp(4,\R)$) and (b) linear couplings
($\OSp(4,\R)$) . The solid lines refer to the control field on the
first qunit, while the dotted refer to the second.}\label{swapf}
\end{figure}

For this gate, we employ the photon model with two local phase
controls, using two different free Hamiltonians: (i) $H_0=x_1p_2$,
which involves squeezing operations; (ii) $H_0=x_1p_2-x_2p_1$,
which involves only linear optics. It can be verified that these
systems are controllable over (i) $\Sp(4,\R)$ and (ii)
$\OSp(4,\R)$, respectively. The simulation results in
Fig.\ref{swapj} show that the optimal search restricted on
$\OSp(2N,\R)$ generally exhibits fast convergence, as in the case
of control of discrete unitary gates. This is not surprising
because the group $\OSp(2N,\R)$ is isomorphic to the unitary group
$\U(N)$ and the corresponding dynamical control system is
equivalent to a $N$-level discrete quantum control system. By
contrast, optimal control using squeezing operators as well
exhibits no advantages compared to using only linear operations;
in fact, the resulting control fields have much greater fluences
(Fig.\ref{swapf}).

\begin{figure}[h]
\centerline{
\includegraphics[width=6in,height=2in]{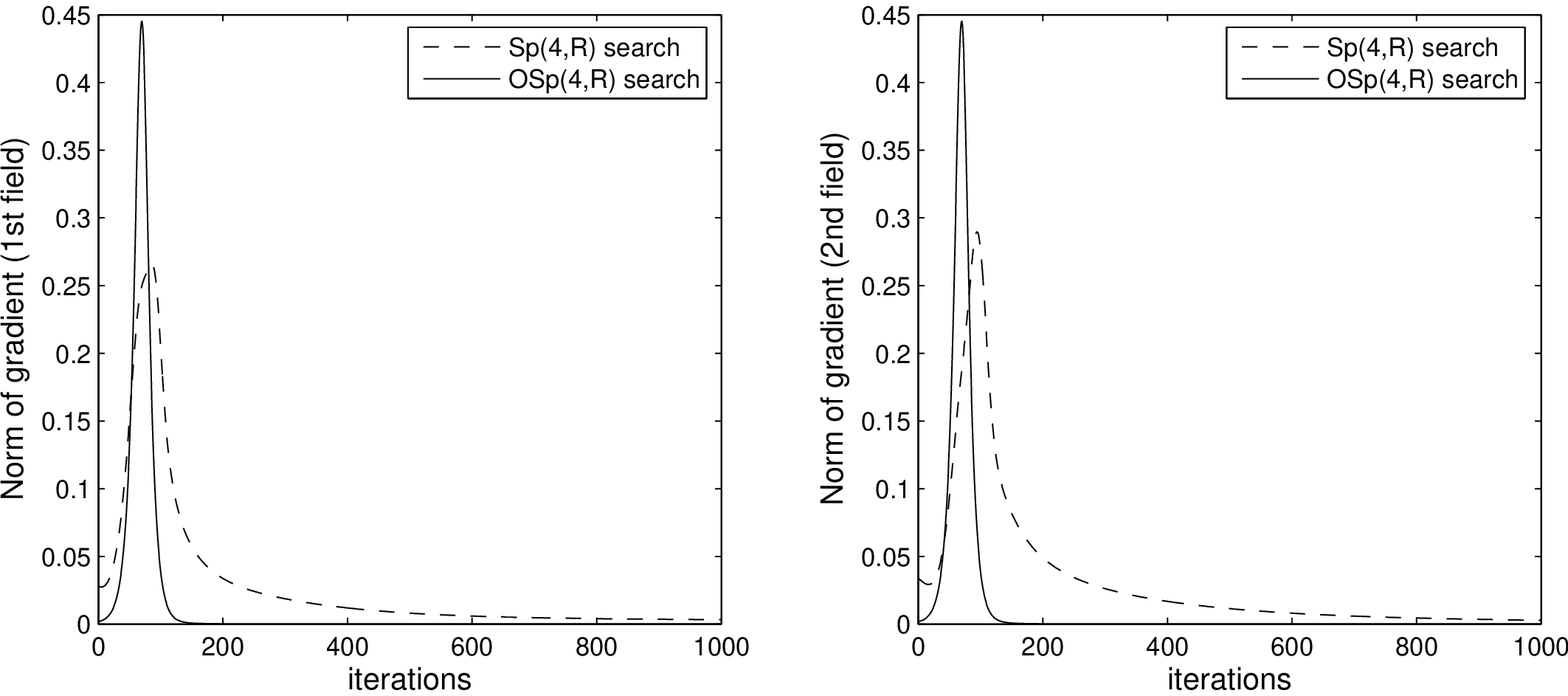}
} \caption{The norm of the gradient vector during optimal search for
the SWAP gate, starting from near a saddle point, with linear and
squeezing couplings ($\Sp(4,R)$) and linear couplings ($\OSp(4,R)$).
The left panel is for the first control field and the right panel is
for the second.}\label{gradientnrm}
\end{figure}

Because $\OSp(2N,\R)$ is isomorphic to $\U(N)$, a comparison of
the gradients of the fidelity on $\Sp(2N,\R)$ and $\OSp(2N,\R)$
sheds light not only on the comparative efficiencies of these two
optimization problems, but also the origin of the slower
convergence of noncompact CV gate optimization versus that of
discrete gate optimization. Fig.\ref{gradientnrm} displays the
norm of the gradient of the fidelity of the SWAP gate with respect
to the control field on $\Sp(2N,\R)$ and $\OSp(2N,\R)$, at each
algorithmic step during the course of optimization. In order to
sample more points along the optimization trajectory, a steepest
descent algorithm was employed in this case, starting from near a
saddle point of the control landscape. As can be seen, the norm of
the gradient is larger on $\OSp(2N,\R)$ at most points along the
optimization trajectory. In addition, over several runs, it was
found that the gradient norm changes more abruptly during
optimization on $\Sp(2N,\R)$ compared to $\OSp(2N,\R)$ (or,
equivalently, $\U(N)$). It was also observed that the components
of the gradient at successive dynamical time points change
abruptly on $\Sp(2N,\R)$. These features, presumably originating
in the noncompactness of $\Sp(2N,\R)$, undoubtedly act to retard
the convergence of searches carried out on this group.

\subsection{Composite algorithms}

Composite algorithms composed of large numbers of low-dimensional
Clifford group gates will generally have richer structures in
their singular values and a greater number of critical manifolds.
The geometry of the control landscape is also expected to be
complexified for such gates. In practice, representing composite
Clifford group algorithms through sequences of 1-qunit or 2-qunit
gates is generally preferred. However, since using a larger number
of gates may increase the likelihood of information loss through
quantum decoherence, the implementation of higher dimensional
transformations is desirable in some instances.

Consider a composite operation on 3 qunits that sums the values of
their $q$-components. This gate can be decomposed into two
elementary gates $\SUM(1,2,3)=\SUM(2,3)\times \SUM(1,2)$, which is
represented by
$$\SUM(1,2,3)=\left(%
\begin{array}{cccccc}
  1 & 0 & 0 & 0 & 0 & 0 \\
  0 & 1 & 0 & 0 & 0 & 0 \\
  0 & 1 & 1 & 0 & 0 & 0 \\
  0 & 0 & 0 & 1 & 0 & 0 \\
  0 & 0 & 0 & 0 & 1 & -1 \\
  0 & 0 & 0 & 0 & 0 & 1 \\
\end{array}%
\right)\left(%
\begin{array}{cccccc}
  1 & 0 & 0 & 0 & 0 & 0 \\
  1 & 1 & 0 & 0 & 0 & 0 \\
  0 & 0 & 1 & 0 & 0 & 0 \\
  0 & 0 & 0 & 1 & -1 & 0 \\
  0 & 0 & 0 & 0 & 1 & 0 \\
  0 & 0 & 0 & 0 & 0 & 1 \\
\end{array}%
\right)=\left(%
\begin{array}{cccccc}
  1 & 0 & 0 & 0 & 0 & 0 \\
  1 & 1 & 0 & 0 & 0 & 0 \\
  1 & 1 & 1 & 0 & 0 & 0 \\
  0 & 0 & 0 & 1 & -1 & 0 \\
  0 & 0 & 0 & 0 & 1 & -1 \\
  0 & 0 & 0 & 0 & 0 & 1 \\
\end{array}%
\right).$$

\begin{figure}
\centerline{
\includegraphics[width=3.5in,height=2.5in]{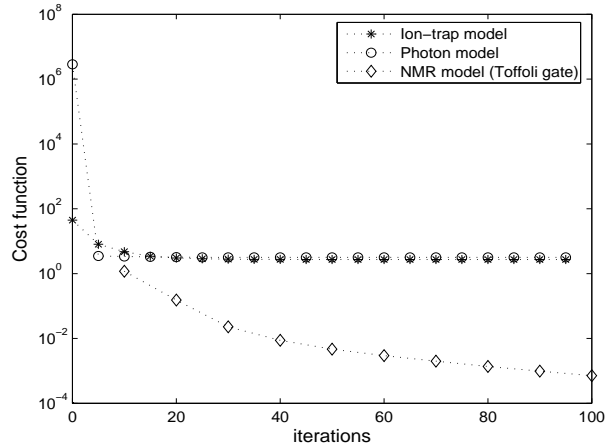}
} \caption{The convergence of optimal searches for the 3-qunit SUM
gate with photon model and ion-trap model, and 3-qubit
Controlled-CNOT gate with NMR model.}\label{3sumj}
\end{figure}

This composite operation can be implemented using a photon model
where the internal Hamiltonian includes interactions between 1-2
and 2-3 qunits, i.e., $H_0=x_1p_2+x_2p_3$, and three local control
Hamiltonians applied in the form of $H_{j}=x_j^2+p_j^2$,
$j=1,2,3$. Again, because the internal Hamiltonian $H_0$ is
noncompact, full controllability is not guaranteed at arbitrary
final time $t_f$. In the case of the ion trap model, the internal
Hamiltonian $H_0$ consists of uncoupled harmonic oscillators, and
two nonlocal controls $$H_{1}=a^\dagger b_1+ab_1^\dagger,\quad
H_{2}=a^\dagger b_2^\dagger+ab_2$$ are applied. As discussed
above, there is no guarantee that the gate will be reachable at
any final time within this model, since the system does not
satisfy the controllability rank condition. In Fig.\ref{3sumj}, we
compare the convergence speeds of optimal control search for the
3-qunit SUM gate using these models with that of its discrete
quantum counterpart, the Controlled-CNOT gate (Toffoli gate). In
the latter case, the NMR control system described above was used.

As can be seen from Fig.10, neither the ion trap nor the photon
control search converges within the specified tolerance for the
chosen final time, whereas the strongly controllable system does
converge. The weakly controllable and uncontrollable systems
therefore display similar behavior for this composite gate. This
example demonstrates that the controllability of a CV gate control
system may become a more important consideration for higher
dimensional gates.

For both discrete and continuous quantum systems, the decrease in
convergence speed with increasing system dimension appears to be
severe; in addition, control field searches are more likely to
become trapped due to easier loss of controllability. Therefore,
for a quantum algorithm involving a polynomially large number of
operations (i.e., primitive gates), it would indeed appear more
efficient to apply sequences of smaller gates rather than
attempting global search over transformations on the whole set of
qunits. The scaling of optimal control search effort with system
size for CV gates is an important subject for future study.

\begin{figure}[h]
\centerline{
\includegraphics[width=6in,height=6in]{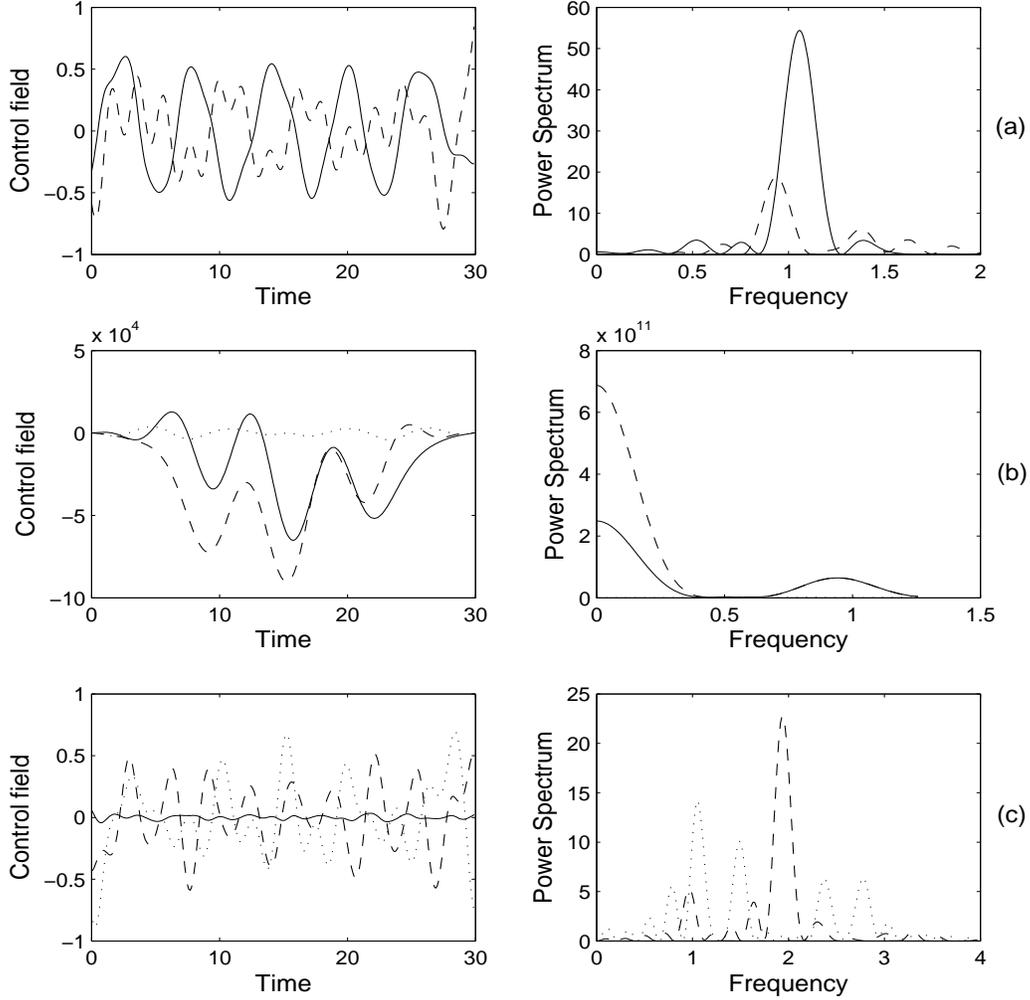}
} \caption{The optimal control fields for
 (a) 3-qunit SUM
gate with ion trap model, (b) 3-qunit SUM gate with photon model
and (c) 3-qubit Controlled-CNOT gate. For (b) and (c), the solid
(dashed, dotted) lines are the control fields on the first
(second, third) qunit/qubit.}\label{3sumf}
\end{figure}

Finally, from Fig.\ref{3sumf}, we observe a stark difference in
the fluence of the optimal control fields obtained using local
versus nonlocal controls. For the photon model (local controls),
the fluence of the optimal fields exceeds physical limits, whereas
for the ion trap model (nonlocal controls) and the NMR model, the
fluence remains bounded. This indicates that for composite gates,
CV control models employing nonlocal controls may be preferable to
those whose design requires the use of local controls. Note that
although the NMR model employs local controls, their fluence
remains small, suggesting that this problem does not arise for
discrete variable gates.

\section{DISCUSSION}

The absence of local traps in control landscapes for symplectic
gate fidelity indicates that given sufficient time, local
gradient-based algorithms will generally succeed at reaching the
global optima (perfect fidelity), assuming the system is
controllable. This property, combined with other attractive
features of continuous QIP such as the high bandwidth of
continuous degrees of freedom, strengthens the feasibility of QIP
over continuous variables.

We have seen that CV gate optimization problems can be divided
into two classes with inherently different complexities.  The
first, wherein only linear Hamiltonians are employed as controls,
is mathematically identical to the problem of discrete unitary
gate optimization. The second, in which squeezing operations are
also employed, is generally more expensive.

A second point of distinction between these two control problems
is the complexity of the optimal fields. The optimal fields for
discrete gate control are typically in resonance with the
transition frequencies of the system, {within the weak-field
regime}. By contrast, in the case of CV gate optimization over
$\Sp(2N,\R)$, the fields are usually not simply related to the
natural resonant frequencies of the system because the Hamiltonian
possesses imaginary eigenvalues. These eigenvalues produce
exponentially increasing and decreasing modes in the field. The
former can result in instabilities during the optimization
process\footnote{In both cases, local controls are generally
associated with optimal fields whose Fourier spectra bear no
simple relationship to system transition frequencies.}. Moreover,
for CV gate optimization over $\Sp(2N,\R)$, there is a strong
dependence of field complexity on the final dynamical time. It is
important to identify several controllable $t_f$'s and choose the
one corresponding to control fields that are physically the most
simple to implement. However, the higher degeneracy of these
fields means that we are presented with more choices for
convenient physical implementation, and point to a rich variety of
distinct control mechanisms that reach the same objective.

Finally, the effects of quantum system controllability on control
optimization are more subtle for CV gate control than for discrete
variable gate control. For CV gate control, it is often difficult
to identify a final dynamical time $t_f$ at which the gate can be
reached with high fidelity, if the system is not strongly
controllable. Moreover, several current physical models for CV
gate synthesis are not fully controllable for any choice of $t_f$.
For these systems, even when the gate of interest is reachable,
the cost of optimal search is typically steep. It is therefore
particularly important to employ physical control systems that
satisfy the conditions for controllability.

These details must be borne in mind when assessing the comparative
difficulty of implementing quantum communication protocols over
continuous versus discrete variables, and underscore the
importance of using optimal control theory in the design of
high-fidelity symplectic gates. An important conclusion of this
work is that it is important to use the methodology of control
theory when designing physical systems for the implementation of
continuous variable quantum gates, both in the choice of
appropriate physical systems and in the determination of the
controls themselves. Application of OCT to practical CV quantum
information tasks will require the imposition of penalty terms on
the control Hamiltonians corresponding to physical constraints.
For control of discrete states in molecular systems, the most
significant constraint on the control fields is the total fluence,
since current pulse shaping technologies are capable of producing
the majority of shapes predicted by OCT. Ongoing efforts to design
pulse shapers with ultra high bandwidth should further facilitate
implementation of the theoretically predicted fields. By contrast,
for CV systems, it is difficult to shape control pulses within
certain physical models. Applying shape constraints to the
optimization problems above would amount to choosing amongst the
highly degenerate sets of control Hamiltonians that solve the
generic OCT problem.

In the context of particular Hamiltonians, time optimal control
theory has been applied to assess the minimal time necessary to
implement various primitive discrete quantum gates
\cite{Khaneja2001,Khaneja2002a,Khaneja2002b}. A natural counterpart
to the current work is the application of time optimal control
theory to symplectic gates, using restricted control Hamiltonians
that are commonly implemented in the quantum optics laboratory.
Given the susceptibility of CV quantum systems to noise, time
optimal control of CV gates is particularly relevant. Such studies
would necessitate determination of the minimal length geodesics in
the noncompact symplectic group, or in subgroups of $\Sp(2N,\R)$
that can be reached using the restricted controls. Preliminary work
\cite{Braunstein2005} along these lines employing the so-called
Bloch-Messiah theorem has been reported, but it is unlikely that
analytical solutions exist for most time optimal symplectic
transformation control problems, especially for higher qunit
systems. As such, the development of numerical OCT algorithms suited
to time-optimal CV gate control is an important future challenge.

In the present work, we have adopted the approach of optimizing a
scalar objective function for gate fidelity. An alternative
approach would be to track a predefined trajectory on the space of
symplectic propagators between the initial and target gate. From a
local perspective, the former is computationally more efficient,
but globally, the latter may offer an advantage. Perhaps more
importantly, the latter approach would provide insight into the
effect of the map between control fields and dynamical propagators
on optimization efficiency. Indeed, the numerical simulations
above indicate that the properties of this map are primarily
responsible for the differences in optimization efficiency between
discrete and CV quantum OCT. Moreover, these properties will
impact the efficiency of control optimization for any observable
of a quadratic CV system, since the symplectic transformation
uniquely defines the infinite-dimensional unitary propagator
corresponding to the CV quantum dynamical process.

Finally, we note that universal quantum computation requires
nonlinear symplectic gates corresponding to the ability to count
photons in the electromagnetic field \cite{LloSlo1998}. The
implementation of nonlinear symplectic gates necessary to achieve
universal continuous quantum computation is known to be difficult
to achieve with high fidelity \cite{GotKit2001}. Purification
protocols are necessary to distill from an initial supply of noisy
nonlinear symplectic states a smaller number of such states with
higher fidelity. Future work should consider the challenges
inherent in implementing such gates through the methodology of
optimal control theory.

\begin{acknowledgments}
The authors acknowledge support from DARPA and NSF.
\end{acknowledgments}


\end{document}